\documentclass[letterpaper]{article} % DO NOT CHANGE THIS
\usepackage{aaai25}  % DO NOT CHANGE THIS
\usepackage{times}  % DO NOT CHANGE THIS
\usepackage{helvet}  % DO NOT CHANGE THIS
\usepackage{courier}  % DO NOT CHANGE THIS
\usepackage[hyphens]{url}  % DO NOT CHANGE THIS
\usepackage{graphicx} % DO NOT CHANGE THIS
\usepackage{natbib}  % DO NOT CHANGE THIS AND DO NOT ADD ANY OPTIONS TO IT
\usepackage{caption} % DO NOT CHANGE THIS AND DO NOT ADD ANY OPTIONS TO IT
\usepackage{tikz}
\def\checkmark{\tikz\fill[scale=0.4](0,.35) -- (.25,0) -- (1,.7) -- (.25,.15) -- cycle;} 

\usepackage{multirow}

\newcommand{\ourterm}{data horizon }
\usepackage{algorithm}
\usepackage{algorithmic}
\usepackage{booktabs}
\usepackage{xcolor}

\usepackage{subfiles}

\usepackage{newfloat}
\usepackage{listings}
\DeclareCaptionStyle{ruled}{labelfont=normalfont,labelsep=colon,strut=off} % DO NOT CHANGE THIS
\floatstyle{ruled}
\newfloat{listing}{tb}{lst}{}
\floatname{listing}{Listing}
\title{Into the Void: Understanding Online Health Information in Low-Web Data Languages}
\author {
    Hellina Hailu Nigatu\textsuperscript{\rm 1}\footnote{corresponding author: hellina\_nigatu@berkeley.edu},
    Nuredin Ali Abdelkadir \textsuperscript{\rm 2,3},
    Fiker Tewelde\textsuperscript{\rm 4}, \\
    Stevie Chancellor \textsuperscript{\rm2}, \
    Daricia Wilkinson \textsuperscript{\rm5}
}
\affiliations {
    \textsuperscript{\rm 1}UC Berkeley \\
    \textsuperscript{\rm 2}University of Minnesota \\
    
    \textsuperscript{\rm 3}Distributed AI Research Institute\\
    \textsuperscript{\rm 4}Addis Ababa University \\
    \textsuperscript{\rm 5}Arizona State University \\
   
}
\begin{document}
\newcommand{\hellina}[1]{\textcolor{orange}{[Hellina Comment: #1]}}
\newcommand{\todo}[1]{\textcolor{red}{[TODO: #1]}} 
\maketitle

\begin{abstract}
% Social media platforms have gradually morphed into informal search engines, with millions of people worldwide relying on them to discover real-time and community-driven information, news, and recommendations. 
% While these platforms offer significant benefits, they also present substantial risks regarding the type of information to which people are exposed---risks that are especially concerning in cases where there is limited data available.
% Search queries for concepts with limited data available lead to data voids--information spaces with limited credible sources, making them easy targets for exploitation by media manipulators. 
% Data voids are prominent in users' interactions with search engines and recommendation systems, particularly for cases where queries have limited information available.
%Prior work has introduced the concept of data voids--information spaces with limited credible resources, making them easy targets for media manipulators. 
%In this paper, we look into 

%As social media platforms increasingly serve as primary search tools for health information seeking, their role in shaping access to reliable health information has grown both influential and problematic with prior work .

Data voids—areas of the internet where reliable information is scarce or absent—pose significant challenges to online health information seeking, particularly for users operating in low-web data languages. These voids are increasingly encountered not on traditional search engines alone, but on social media platforms, which have gradually morphed into informal search engines for millions of people. In this paper, we introduce the phenomenon of \textit{data horizons}: a critical boundary where algorithmic structures begin to degrade the relevance and reliability of search results. Unlike the core of a data void, which is often exploited by bad actors to spread misinformation, the data horizon marks the critical space where systemic factors, such as linguistic under-representation, algorithmic amplification, and socio-cultural mismatch, create conditions of informational instability. Focusing on Tigrinya and Amharic as languages of study, we evaluate (1) the common characteristics of search results for health queries, (2) the quality and credibility of health information, and (3) characteristics of search results that diverge from their queries. 
We find that search results for health queries in low-web data languages may not always be in the language of search and may be dominated by nutritional and religious advice. 
We show that search results that diverge from their queries in low-resourced languages are due to algorithmic failures, (un)intentional manipulation, or active manipulation by content creators. 
We use our findings to illustrate how a data horizon manifests under several interacting constraints on information availability.

\end{abstract}

\begin{links}
    \link{Supplementary Material}{https://osf.io/dk7zc/}
    % \link{Datasets}{https://aaai.org/example/datasets}
    % \link{Extended version}{https://aaai.org/example/extended-version}
\end{links}
\section{1 Introduction} \label{sec:intro}

% \begin{quote}
%     \textit{The enjoyment of the highest attainable standard of health is one of the fundamental rights of every human being without distinction of race, religion, political belief, economic or social condition.} --WHO
% \end{quote}

    % Situation: What is the current state of the situation.
% social media source of health information
% general information not avaible online
% Health information even more lacking
% However, medical content is not easily available in all languages. Particularly for low-web data languages--where, by definition, the languages lack data on the web \cite{}--there is limited health information available~\cite{}. 

Social media platforms such as YouTube and TikTok are used as sources of information, including in critical settings like healthcare \cite{ozdemir_metric_2021}. Users rely on these platforms to gain general information on health-related concerns~\cite{chirumamilla_patient_2021}, to seek support from others who may have similar health conditions~\cite{naslund_naturally_2014, freeman_how_2023}, or to learn more about planned medical procedures~\cite{roberts_social_2023}. As such, these platforms have gradually morphed into important sources for health-related information. 

Certain features of social media platforms make them appealing channels for health communication. 
For health information providers, social media platforms offer instant access to a wide range of audiences~\cite{jafar_social_2023}. This quick and broad reach of social media platforms makes them especially attractive in public health crises such as pandemics~\cite{talie_fenta_role_2024}. 
For health information seekers, social media platforms provide a unique space for user-generated medical content that is ``relatable'' and easier to digest for the general public~\cite{milton_i_2023}. However, while these affordances benefit the health information landscape, they could also be a source of harm. The far and immediate reach of social media platforms means that health misinformation can spread just--if not more--as quickly and as widely as credible health information~\cite{sylvia_chou_where_2020}. Further, while users value the relatability of social media health information, they at times find personal anecdotes untrustworthy~\cite{quintero_johnson_optimizing_2017}.

% \textcolor{red}{DW: I think the overall argument needs to be strengthened and be better connected. Maybe starting out how it is now by speaking about the prevalence of Youtube and TikTok as sources of health information. Explaining the motivation behind why people use these platforms and making an argument about the roles of language disparities and its contribution to health information access or the lackthereof. Then you transition to data voids and explain how this is made to be more complex as a sociatechnical issue being intertwined with social and infrastructural challenges in healthcare access locally.}

    % Implication: What’s the consequence of failing to act on the problem or opportunities you described in the Complication?
% data void
% human manipulation
% algorithmic manipulation
% models trained on this data.
% Health information is not equally available in all languages and to all communities.
% Socio-economic fa/ctors influence whether or not communities have access to online health information.
In the complex health information seeking landscape, language is a critical component affecting how health information is retrieved, distributed, and consumed.
%one determining factor in whether or not individuals have access to online health information.  
% One such factor is language.
As \citeauthor{nigatu_zenos_2024} argue, languages--and their speakers--may lack economic, human, data, and technological resources, creating a disparity in what language technologies are available to the speakers.  
For instance, most African languages are \textit{low-web data languages}--i.e, they have limited digitally available data, which inherently limits the online health information available for their speakers~\cite{hu_natural_2025}. 
The lack of information in a critical setting like healthcare is arguably a form of harm by itself, as it creates a disparity in health information access~\cite{ukonu_socio-cultural_2021}. 
The health information gap means that speakers of these languages may be unable to access necessary--and at times life-saving information~\cite{al_shamsi_implications_2020}. 
This problem is especially precarious in African communities, where access to healthcare services is limited~\cite{oleribe_identifying_2019}.
% Access to quality health information is also linked to overall improvement in quality of life~\cite{}. 
Further, healthcare information available to these communities might be dominated by content from unverified sources. 
Prior work has shown that, despite high demand for natural cures by African users, search results provide limited high-quality resources on natural remedies~\cite{abebe_using_2019}.
% \citet{abebe_using_2019} studied information seeking patterns in 54 African countries and found limited high-quality health information for natural cures, even though natural cures based on herbal medicines were sought after, especially by users of older age groups. 
% Thus, health queries in languages with limited digital resources may lead to low-quality, unverified information that poses harm to individuals' well-being. 

In this paper, we address a growing concern surrounding online information seeking: the proliferation of data voids. \citeauthor{golebiewski_data_2019} define data voids as ``voids [that] occur when obscure search queries have few results associated with them''~\cite{golebiewski_data_2019}. In low-web data languages, data voids are not merely gaps in search results but pose significant risks in critical settings like health information, where poor search results can have severe consequences. To understand how platform design, language, and sociocultural context intersect to create or exacerbate information breakdowns, we posit the following questions:

\begin{itemize}
    \item \textbf{RQ1:} What are the most common characteristics of search results retrieved for health-related queries in low web-data languages
    
    \item  \textbf{RQ2:} What is the quality of health information retrieved for low-web data languages? 
    
    \item  \textbf{RQ3:} When search queries and results do not align, what are the common characteristics of those results?
\end{itemize}

%Action: What do you want your audience to do, believe, or understand as they are listening to you?

%%%%%%%%%%%%
%Data voids are (1) exploited by media manipulators and users with adversarial agendas, and (2) surfaced when there is an external trigger that increases traffic to a particular search query~\cite{golebiewski_data_2019}.
Based on our findings, we introduce the concept of \textit{data horizons}--the threshold where the quality and relevance of search sharply diminishes, marking the boundary of the start of a data void and information collapse.
% --where algorithmic structures disintegrate
A data horizon is created from tensions among systemic constraints on information access, disintegrating the fabric of the information landscape into a potential data void. Contrary to a data void where external factors shape whether the void surfaces, at the data horizon, users are exposed to problematic content due to inherent limits in systemic structures. It is a tipping point in the user experience, beyond which the quality and credibility of information rapidly decline, and the space becomes more vulnerable to manipulation by nefarious actors. 
%As described in the previous paragraphs, health information seeking in languages with limited digital resources is one such case with systemic constraints on information access. We explore how these constraints lead to a data horizon by answering these research questions:

We find that search results for low-web data health queries are dominated by natural remedies and religious content. 
We also find that most of the health information is produced by individual medical professionals whose credentials are not always easy to verify. 
Our findings also show that the health information gap in low-web data languages is exacerbated 1) algorithmically due to the inherent limits of information in the low-web data languages, (2) (un)intentionally through the spread of unverified remedies, even for chronic diseases with known interventions, and 3) by content creators who use non-medical terms to boost their content.
% We contextualize our findings through the broader literature of epistemic injustice\cite{}, and show how the interacting social, political, and economic constraints posed on low-web data language speakers result in an \textit{epistemic tension}\cite{}. 
% Benefit: How will taking your suggested recommended actions affect the audience? What will the outcome(s) be?
Our contributions can be summarized as follows: 

% \textcolor{red}{@Hellina: remember to add section labels}

\begin{itemize}
    \item We provide qualitative and quantitative insights into characteristics of search results for health-related queries in low-web data languages (Section 4.1) and the quality of health information that is available (Section 4.2). 

    \item We identify 5 sources for search results that diverge from their queries when seeking health information in low-web data languages (Section 4.3).
    
    \item We introduce the concept of \textit{data horizons}--boundaries right before a data void, where the information fabric disintegrates as a result of the tension caused by systemic constraints in information access (Section 5).

     \item Based on our findings, we provide recommendations for stakeholders in protecting individuals from crossing the data horizon into the void (Section 6). 
    
\end{itemize}

% Our work contributes a framew
% The remainder of this paper is organized as follows: In Section \ref{sec:background}, we provide background on our languages of study and summarize related work in access to healthcare and healthcare information for African communities. In Section \ref{sec:rq1}, we detail our methods for answering \textbf{RQ1} and provide our findings on types of health information available in our languages of study. In Section \ref{sec:rq2}, we detail the methods we used to answer \textbf{RQ2} and provide our findings on who shares health information in these languages. Then in Section \ref{sec:rq3}, we go present our findings to answer \textbf{RQ3}. We discuss the limitations of our study in Section \ref{sec:limitations} and provide our recommendations and reflection of our findings in Section \ref{sec:discussion}. We conclude in Section \ref{sec:conclusion}. 

\section{2 Background and Related Work} \label{sec:background}
% \textcolor{red}{DW: this section should be reorganized and just have a related works. First, have a post sign (2-3 sentences describing how the section is organized. Provide more background on health communication/health information seeking and the role of social media, then discuss the importance for low web data languages and African users, then you could justify the scope with the languages used. The end of each sub-section should describe limitations in existing knowledge and how our work fills this gap}
% \section{Related Work} \label{sec:related_work}
% Here, we contextualize our study in related work and provide background for the scope of our study. First, 
In this section, we first provide a literature review on health communication on social media platforms. Then, we discuss the current landscape of online health information for low-web data language speakers. Finally, we present information on our two languages of study. 

\subsection{2.1 Health Communication and Social Media}
% \hellina{Make sure we are NOT making any health claims. It should be clear we are not healthcare professionals.}
Health communication refers to ``the study and use of methods to inform and influence individual and community decisions that enhance health.''\cite{freimuth_contributions_2004}. 
For health communication to be effective, it has to be (1) understandable by the general public, requiring careful interpretation of dense technical and research concepts~\cite{dahm_tales_2012}, (2) culturally relevant and contextual~\cite{brooks_culturally_2019, tang_who_2024}, and (3) transmitted through channels of communication that are widely accessible by the community~\cite{lee_communication_2020}.
% Disseminating health information to the general public is complicated by the need to translate dense technical and research concepts to formats understandable and applicable to the general public \cite{}.
% Additionally, effective dissemination requires an understanding of the socio-economic and cultural backgrounds of the audience \cite{}. 
% According to the WHO\cite{}, ``high-quality health information must be understandable, engaging, and culturally competent to be influential.'' 
% To ensure the information reaches the right audience, one also needs to identify an effective channel of communication\cite{}. 
% One of the principles of the WHO in their framework for effective communication of health information is accessibility\cite{}. 

With the pervasive use of social media, platforms like YouTube and Facebook have opened avenues to reach a wider set of audiences across geographies~\cite{jeyaraman_multifaceted_2023}. 
However, social media platforms have also opened the door for widespread misinformation, which in the context of healthcare information could lead to direct physical and mental harm~\cite{ghenai_fake_2018}. Usually, such misinformation is disseminated by non-medical professionals~\cite{bautista_healthcare_2021}. Yet, getting accreditation on social media platforms as a credible health information content creator is not an easily available process for all medical professionals. For instance, YouTube has a set of criteria for content creators to be registered as a source of health information on the platform~\cite{youtube_apply_2021}. Currently, the criteria and processes are available for healthcare providers from 10 countries\footnote{The countries are USA, UK, Germany, Japan, India, Indonesia, Canada, South Korea, France, Brazil, and Mexico~\cite{youtube_apply_2021}}.; for healthcare providers outside those geographic locations ``organizations with pre-existing, standard vetting mechanisms such as accredited hospitals, academic medical institutions, public health departments, or government organization [...] can submit a request.''~\cite{youtube_apply_2021} However, the request may take ``months to show up in our [YouTube's] features.''~\cite{youtube_apply_2021}.
% As a result, communities from certain countries may not have access to verified health information on social media platforms. 
Hence, depending on their location, some communities may have a hard time accessing contextual, verified health information.

The challenges in distributing verified health information result in a health information gap--where quality and verified health information is (1) not available or (2) not accessible to certain communities. Despite this gap, individuals increasingly rely on online platforms for their health information needs~\cite{zhang_pragmatic_2023}. In addition to their accessibility, social media platforms create a space where users find ``relatable'' health information through shared personal experiences~\cite{milton_i_2023}. However, not all health information needs are met through personal stories: for instance, some creators rely on (pseudo)anonymity to reduce risks associated with the self-disclosure of sensitive information while some users may avoid consuming health information based on personal experiences due to the anonymity of content creators making it difficult to verify the credibility~\cite{milton_seeking_2024}. As a result, there is a tension where users require authoritative, verified health information and at the same time seek ``relatable'' content on social media platforms~\cite{milton_i_2023, viviani_credibility_2017}. Further, online health information allows for seeking information on non-conventional medical information--for instance, users may seek and consume information on traditional, complementary, and integrative medicine (TCIM)~\cite{trubner_health_2025}. Although such information may be useful and perceived as more reliable and accessible in some contexts, prior work has also raised concerns for how limited instructions and credible sources for information could create (un)intended risks for health information seekers~\cite{karusala_towards_2022}. 

% In addition to the repeatability and social aspect of health information on social media platforms, users may also seek health content online due to lack of access to physical medical care 

% \subsection{Health Information Gap}

\subsection{2.2 Health Information and African Communities}
 While the health information gap is a global phenomenon, it does not affect all communities equally. Several studies have demonstrated that marginalized groups are more likely to seek health information online~\cite{ukonu_socio-cultural_2021, nangsangna_factors_2019}. Yet, there is limited online health information available for such communities.
 
 Particularly for African communities, access to quality online health information is limited due to multiple interacting factors: 
 % anguage barriers, lack of culturally and socially relevant materials, resource limitations to verified digital resources. 
 First, as discussed in Section 1, African languages are often categorized as low-web data languages whose lack of digital data is further exacerbated in the context of healthcare~\cite{hu_natural_2025}. 
 Secondly, online health information that is available may not account for the cultural and social backgrounds of African communities~\cite{usman_online_2024}.
 Thirdly, there is a lack of verified health information from authoritative institutions for African communities~\cite{abebe_using_2019}. 

Prior works show that health information seeking behaviors are affected by the socio-cultural backgrounds of information seekers. In a large-scale analysis of search queries from the 54 African countries, \citet{abebe_using_2019} found that women were more likely to search for content about pregnancy and breastfeeding, while men were more likely to search for ``cure news.'' Socio-cultural backgrounds may also influence the medium through which users access online health information:  \citet{ukonu_socio-cultural_2021} found that women in Southeast Nigeria (financially) depended on their spouses for avenues to access health information. \citet{njenga_issues_2024} found that the elderly in Kenya relied on relatives to seek online health information on their behalf as the platforms were inaccessible to them.
% \hellina{types of health info African c seek}

% \hellina{Bad outcomes}

\subsection{2.3 Study Context}
As we have summarized in the sections above, search queries may surface irrelevant or unverified information for speakers of low-web data languages. 
This increases the probability of low-web data language health queries returning ``obscure'' results, thus, intensifying the health information gap into a potential data void.
\citet{golebiewski_data_2019} showed that in the context of health, media manipulators posting anti-vaccination conspiracy content associated their content with other videos that were health-related. In the context of high-web data languages like English and socio-cultural context of the US, \citet{golebiewski_data_2019} state that algorithmically ``government and health professional content is more highly weighted for search queries related to vaccination.'' 
However, in the context of low-web data language speakers, verified health information from authoritative institutions is not always available, thereby, affecting the ranking of available content. 

While prior work has explored online health information seeking behavior, or data voids, or the impact of socio-cultural aspects in more of an isolated capacity, there is an urgent need to investigate the intersection of this trio. In this study, we address this gap and carefully examine the anatomy of this space, revealing a new phenomenon that has not been explored--a \ourterm. 
We explore this phenomenon by collecting data for health queries in two low-web data languages: Amharic and Tigrinya. 
% \textcolor{red}{DW: needs a transition sentence here to connect with previous points or to justify/motivate/rationalize the scope}
% We focus our study on two languages: Amharic and Tigrinya. 
% Both languages are Afro-Semitic languages and use the Ge'ez script. 
Amharic is spoken in Ethiopia by an estimated 57.5 million people worldwide~\cite{basha_detection_2023}. Tigrinya is spoken in Ethiopia and Eritrea by an estimated 10 million people worldwide~\cite{haile_error_2023}. Both languages are Afro-Semitic languages and use the Ge'ez writing script, which is an Abugida writing system--each character in the script represents a consonant and a vowel\footnote{https://www.omniglot.com/writing/ethiopic.htm}.

We selected YouTube and TikTok as platforms of study. Prior work has demonstrated that TikTok has garnered more views as a source of information for certain health conditions~\cite{alkhodair_assessing_2024}. 
Prior research has also explored the use of YouTube as a source of health information and found that the majority of the content is not posted by medical professionals~\cite{clarke_polycystic_2023}. In online health information literature, the two platforms also present a contrasting case for video information sharing, in length and quality of video, where TikTok has shorter, lower quality videos~\cite{munoz_youtube_2024}.

\section{3 Methods} \label{sec:methods}
We used a mixed-methods approach to answer our research questions: we used topic modeling to automatically extract salient topics from our collected data and used qualitative analysis to interpret the patterns in our data. In this section, we first describe how we collected our dataset and give details on each of our methods.

\subsection{3.1 Dataset Preparation} \label{sec:dataset}
Our study goal is to understand the characteristics of the health information gap on YouTube and TikTok in two low-web data languages. Below, we first describe how we prepared queries in the two low-web data languages. Then we describe how we collected data from the two platforms using the queries we prepared. 
% the respective APIs for each platform to collect our data. Below, we first describe how we prepared our queries

% Below, we give the details of our data preparation(\textsection 4.1) and analysis(\textsection 4.2).

\begin{table}[]
    \centering
    % \small
    \footnotesize
    \begin{tabular}{p{2.8cm}|p{3.5cm}|p{.7cm}}
    \toprule
        \textbf{Disease Category}  & \textbf{Examples} & \textbf{Total Count} \\
        \midrule
        Communicable  & Meningitis, Tuberculosis & 6\\ 
        Non-Communicable  & Diabetes, Cancer& 7\\ 
        Reproductive Health  & Chlamydia, Gonorrhea & 5\\ 
        Women and Children  & Cervical cancer, Measles & 11\\ 
        Mental Health  & Anxiety, Depression& 4\\ 
        Other & Vaccination, Malnutrition  & 5\\
        \bottomrule
    \end{tabular}
    \caption{Health conditions used to create search queries in the two languages of study.}
    \label{tab:health_conditions}
\end{table}

\paragraph{Preparing Queries} We first collected diseases and health conditions, prioritizing health conditions most prevalent in the Ethiopian and Eritrean communities according to the World Health Organization (WHO), Center for Disease Control (CDC), and the Ministry of Health of Ethiopia (MOH). This resulted in 35 diseases across 5 categories. 
Table \ref{tab:health_conditions} shows the five disease categories along with the number of diseases in each category and some example health conditions. 
For YouTube, the first and second authors, who are native speakers of Amharic and Tigrinya, respectively, prepared queries in each language separately from the list of diseases and health conditions. After preparing the queries independently for each language, the authors discussed the queries to understand the similarities and differences in the prepared queries across the languages. We focused on four themes in the list of queries we prepared: prevention, symptom, treatment, and cure.  
For each health condition, we prepared 2-4 queries to account for spelling differences, paraphrases, and alternative disease names\footnote{For instance, some diseases are referred to in their English name, transliterated to the Ge'ez script, and with their local language name. Hence, we prepare separate queries with the transliterated and the local language name.}. We refined our queries in weekly meetings where we discussed potential ambiguities and resolved confusions or disagreements. By the end of our process, we had 219 queries for Amharic and 246 queries for Tigrinya for YouTube. 
For TikTok, we used the names of the diseases and health conditions instead of the full question queries, as the TikTok API returns search results for individual keywords. As a result, we did not have four stages of disease progression for TikTok. \footnote{More details on query preparation and sample queries can be found in the Supplementary Material.} 

\paragraph{Collecting Data} We used the \texttt{YT-DLP} API\footnote{https://github.com/yt-dlp/yt-dlp} to collect data from YouTube and the \texttt{TikTok Research Tools} API\footnote{https://developers.tiktok.com/doc/about-research-api} to collect data from TikTok. For YouTube, we first collected the top 10 search results for each of our queries. We then collected channel information for videos returned for our search results. Our data collection resulted in 1560 unique search results and 728 channels. Since the TikTok API only allowed for a maximum of 30 days when indicating time range, we focused our data collection on the most recent 3 months. In total, we collected 397 unique videos from October 2024 to December 2024 from TikTok. Since the TikTok API required us to specify the location of the content creators, we collected data from the same 5 locations in \cite{nigatu_i_2024}: United States of America, United Arab Emirates, United Kingdom, Saudi Arabia, and Ethiopia. Prior work indicates that these countries are best-suited for studying online activity of Ethiopia, Eritrea, and the diaspora~\cite{nigatu_i_2024}.  

\subsection{3.2 Analysis Methods}
As stated above, we used a mixed methods approach relying on both automated and human analysis to answer our research questions. Below, we give details on the three methods we used to analyze our data. 

\paragraph{Topic Modeling} To understand what topics emerge from our collected search results, we trained Latent Dirichlet Allocation (LDA) \cite{blei_latent_2003} models.
We set the number of topics to correspond to the number of diseases in each of the disease categories to identify the topics that are salient in each of the categories(see Table \ref{tab:health_conditions}). We trained models for the video titles and descriptions, and identified the salient terms in our data.  

\paragraph{Content Analysis} 
% \textcolor{red}{DW: the actual method is content analysis and you used ti CRAP framework as a tool to assess credibility and thematically analyze the corpus}
We conducted two qualitative content analyses on the data.
First, to assess the credibility of the health information available in the two low-web data languages, we used the CRAP framework~\cite{beestrum_molly_crap_2007}. The CRAP framework is recommended by the WHO and the National Academy of Medicine to evaluate the credibility of health information online~\cite{raynard_s_kington_identifying_2021}. The CRAP framework has four axis for evaluation: \underline{C}urrency, \underline{R}eliability, \underline{A}uthority and \underline{P}urpose. We used the guiding questions from \cite{raynard_s_kington_identifying_2021} to perform our analysis\footnote{Guiding questions can be found in Supplementary Material.}. 
For this analysis, we first subsampled our dataset for qualitative analysis. For YouTube, we selected the most viewed videos for each disease category (n=5) and each disease progression stage (n=4). 
% For instance, we select the most viewed video for Communicable-Prevention, Communicable-Transmission, and so on.
With this, we had 20 videos per language for a total of 40 videos across the two target languages. For TikTok, we selected the most viewed video in each disease category (n=5), resulting in 10 videos across the two languages of study. This resulted in a total of 50 videos.  
We then used content analysis to understand the common characteristics of search results and to identify the cause for search results that diverge from their queries. 

The first, second, and third authors of the paper conducted the qualitative coding for both analyses.
The first author is a native Amharic speaker (L1) and has basic conversational skills in Tigrinya (L3). The second author is a native Tigrinya speaker (L1) and is a fluent Amharic speaker (L2). The third author is a native Amharic speaker (L1). The authors divided the videos among themselves and analyzed each video by opening and viewing the content. 
The authors first met to review an example video--which was randomly sampled outside of the 50 videos--and discussed the application of the CRAP framework to the sample video. This allowed the authors to resolve ambiguities about the framework. The authors then independently applied the CRAP framework to their assigned videos and met to discuss their observations. Once the authors went through the first round of annotations, they met with the last author to iteratively discuss and refine the themes that emerged. We did not calculate inter-rater reliability as each video was annotated by a single person; instead, we relied on the iterative nature of our analysis to reduce disagreements. Finally, the first and second authors wrote up the final findings together, going over each theme identified and discussing general observations.

% The goal of \textbf{RQ2}, is to understand who is creating health information in the two target languages. We first looked at the statistics of channels and accounts that posted videos in our dataset. We then qualitatively analyzed a subset of the videos to get a deeper understanding of the content creators.
% \hellina{Is there a guideline on steps for health info distribution? }
% \paragraph{Queries}

% \begin{table*}[]
%     \centering
%     \begin{tabular}{c|c|c|c}
%     \toprule
%         Disease & Query Category & Amharic 
%         Queries & Tigrinya Queries \\
%         \midrule
%         \multirow{3}{*}{Malaria} & PREVENTION &  & \\
%         & SYMPTOM & &  \\
%         & TREATMENT & & \\
%         & CURE & & \\
%         \midrule
%         \multirow{3}{*}{HIV} & PREVENTION &  & \\
%         & SYMPTOM & &  \\
%         & TREATMENT & & \\
%         & CURE & & \\
        
%     \end{tabular}
%     \caption{Caption}
%     \label{tab:my_label}
% \end{table*}

% Qual Analysis -- To partition
% Stevie's Paper -- 1 paragraph

% Returned health queries <--> 

% \begin{figure}
%     \centering
%     \includegraphics[width=0.5\linewidth]{figure1.pdf}
%     \caption{Caption}
%     \label{fig:disease_progression}
% \end{figure}
% \subsection{Automatic Analysis}
% \TODO{What automatic Methods can we use here? Topic Modeling? }
% \TODO{Top Viewed; Top Liked; Verified Channels}

\section{4 Results}
In this section, we present the findings from our analysis. We organize this section according to our research questions: we first answer the common characteristics in content retrieved for health queries in low-web data languages. We then present our analysis of the quality of health information using the CRAP~\cite{beestrum_molly_crap_2007} framework. Finally, we give insights into the characteristics of retrieved content that diverges from the search queries.

\subsection{RQ1: What are the most common characteristics of search results retrieved for health-related queries in low web-resourced languages?} \label{sec:rq1}
% \textcolor{red}{DW: Consider rephrasing this RQ. It reads a bit vague. Not exactly clear what "themes" really means here. Option: What are the most common characteristics of health information YouTube and TikTok content retrieved in low web-resourced languages? OR What health-related terms and topics are most salient in YouTube and TikTok content retrieved in low web-resourced languages? OR What health-related terms and topics are most salient in YouTube and TikTok content retrieved in low web-resourced languages and what are the common characteristics? }
% Health information gaps exist when there is (1) no content in a given language or (2) when the information available is not accurate or coming from a credible source~\cite{}. 
% Our \textbf{RQ1} looks into what themes appear in search results for health-related queries in low-web data languages on online platforms. The goal of \textbf{RQ1} is to get a better understanding of what the content available in low-web data languages is in the context of healthcare. 

% \textcolor{red}{DW: a bit repetitive. Perhaps consider making the language clearer}
% \paragraph{Summary} 
We identified two high-level characteristics in search results for health queries in the two low-web data languages: (1) language mixing--where the returned results for some health queries are in languages other than the language of the search, and (2) nutritional advice and religious content returned for health-related queries. Across the two platforms, we found that TikTok results included more specific advice (e.g. videos referring to particular nutritional items as a solution) as compared to YouTube, where the videos were more general. Overall, we find that the search results may not address the health information needs of speakers of low-web data languages. Below, we go into the details of these findings. 

% We found that search results returned for health-related queries show patterns of language mixing--where the returned videos are not in the language of the search. Additionally, we observed patterns of nutritional advice and religious content returned for health-related queries in both languages. Comparing the two platforms, we note that TikTok had more specific salient terms, including referring to particular nutritional items (e.g. `ginger') as compared to YouTube, which had more generic salient terms. Further, we qualitatively identified patterns in search queries and search results where the search result diverges from the health query. 

% Overall, we find that the health information available in the two low-web data languages (1) maybe not be available in the language of search, (2) may not be accurate--i.e may be about an unrelated health condition or not be health related. Below, we go into detail on each of the findings. 

% \hellina{should i spell out here that this connects to how the literature defines health info gap or is it clear?}

\paragraph{Language Mixing:} We find that searching for health-related queries in low-web data languages may surface search results in languages other than the language of the search. 
% In some cases, the entire content 
From our topic modeling analysis, we find that salient terms for Amharic and Tigrinya search results for YouTube include Arabic and English terms (Figure \ref{fig:salient_terms}). 
Qualitatively, we find that Arabic videos in our data were about Reproductive Health and STIs: for instance, searching for symptoms of chlamydia in Amharic returned an Arabic YouTube video about the disease; searching for symptoms of measles in Tigrinya returned an Arabic video about gonorrhea. For language mixing with English, we find that some videos that were in the language of search might have their titles in English in addition to the language of search:  for instance, a video about symptoms and treatments for cancer in Tigrinya had the title in both Tigrinya and English. For Tigrinya search queries, we also observed that some of the videos returned are in Amharic: for instance, the search result for a Tigrinya query about meningitis included videos about the disease in Amharic. As a result, a person searching in a low-web data language may not be able to access the health information they seek, as it is not available in their language.

\paragraph{Salient Topics:} We find that videos in the two languages mostly included nutritional advice and natural remedies. This is evident in the salient terms from our topic modeling (Figure \ref{fig:salient_terms}), where our analysis reveals that terms like \texttt{`foods'}, \texttt{`ginger'}, and \texttt{`rosemary'} appear as salient terms across the two platforms. We also find that religious content appears in the list of salient terms, more dominantly for Amharic search results on TikTok (Figure \ref{fig:salient_terms}). Qualitatively, we found that religious content is primarily returned for mental health-related queries, ranging from teachings by religious leaders to prayer videos targeted towards ``relieving stress'' without support for professional intervention or suggestions for available resources.  
For instance, the query ``what are medications for anxiety?'' returned a 1 hour and 34 minute prayer video on YouTube with 1.8 million views. This suggests that there might be a tendency among consumers of health information in low-web data languages to engage with religious content for mental health-related topics. 

\paragraph{Comparison of Search Results Across Platforms:}
% To get a better understanding of the differences and similarities between the two platforms in search results for health related queries in low-web data languages, we compared the search results we collected from the two platforms. 
Search results from both platforms were similar in terms of including videos that were in languages other than the language of the search. 
We also observed that search results across both platforms included nutritional advice and religious content. 
% For both platforms, we observe salient terms that suggest content that has nutritional advice and natural remedies.
However, we find that the salient terms in the results returned for TikTok tend to be more granular-- as Figure \ref{fig:salient_terms}, shows we observe more specific food and natural remedy terms like \texttt{`ginger'}, \texttt{`rosemary'}, and {\texttt{`lemon juice'} in both Tigrinya and Amharic when queries were conducted across the five disease categories whereas the salient terms on YouTube were more generic such as \texttt{`food'} more broadly.
We also observed spiritual and religious terms like \texttt{`devil'}, \texttt{`evil spell'}, \texttt{`Islam'}, and \texttt{`God'} for communicable, non-communicable, and reproductive disease categories in Amharic on the TikTok search results. Further, as Figure \ref{fig:salient_terms}\footnote{We put a screenshot of the Table due to LaTeX restrictions on using Ge'ez characters.}  shows, TikTok results had more salient terms that were not in the language of search; this is especially more pronounced for Tigrinya, where the salient terms for reproductive health, women and children, and mental health queries included Amharic words.

% \paragraph{Where the Health Information Gap is }
% For mental health, we observe `drug' in both Amharic and Tigrinya and `sleep' in Amharic.  

% % \usepackage{multirow}
% \begin{table}[]
% \begin{tabular}{|l|ll|}
% \hline
% \multirow{2}{*}{Disease Category} & \multicolumn{2}{l|}{YouTube Salient Terms} \\ \cline{2-3} 
%                                   & \multicolumn{1}{l|}{Amharic}   & English   \\ \hline
% Communicable                      & \multicolumn{1}{l|}{  \begin{geez}\font\zz="[AbyssinicaSIL-Regular.ttf]" at 7pt \zz 
% ትርፍራፊውን ማጽዳት ግን ይቀጥል።  \end{geez}}          &           \\ \hline
% Non-Communicable                  & \multicolumn{1}{l|}{}          &           \\ \hline
% Reproductive Health               & \multicolumn{1}{l|}{}          &           \\ \hline
% Women and Children                & \multicolumn{1}{l|}{}          &           \\ \hline
% Mental Health                     & \multicolumn{1}{l|}{}          &           \\ \hline
% \end{tabular}
% \end{table}

\begin{figure*}[t]
    \centering
    \includegraphics[width=\linewidth]{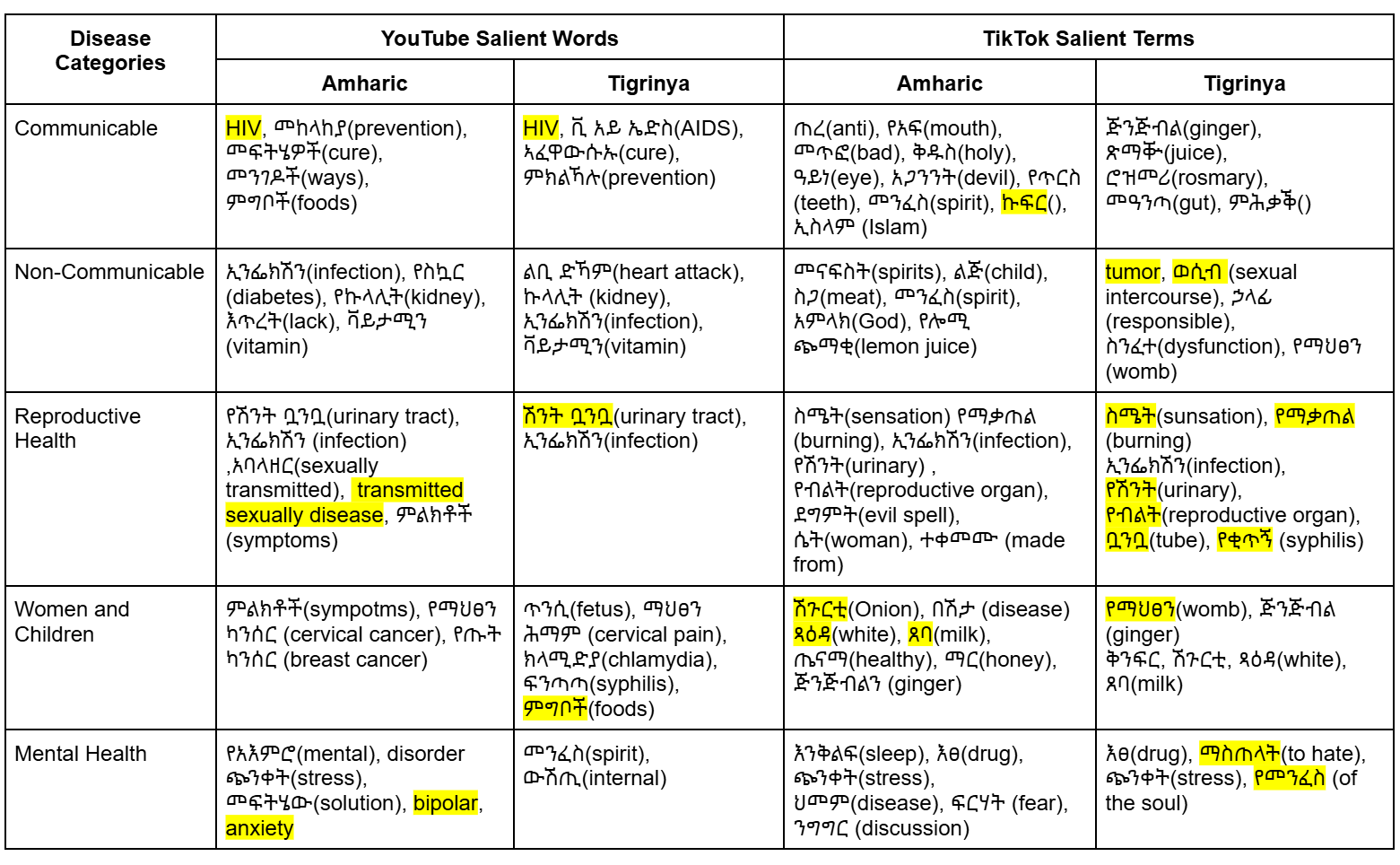}
    \caption{Topic modeling results demonstrating salient words for both platforms (YouTube and TikTok) across different disease categories in the two studied languages (i.e. Amharic and Tigrinya). Terms that are not in the search language are highlighted in \colorbox{yellow}{yellow}.}
    \label{fig:salient_terms}
\end{figure*}

% \begin{figure}
%     \centering
%     \includegraphics[width=\linewidth]{images/tables/tiktok_salient_category.png}
%     \caption{Caption}
%     \label{fig:tiktok_salient_category}
% \end{figure}
% \paragraph{Salient Terms by Disease Progression Stage:}

% \begin{table}[]
%     \centering
%     \begin{tabular}{c|c|c}
%         Stage & YouTube Salient Words & TikTok Salient Words \\
%          Prevention & \begin{geez}\font\zz="[AbyssinicaSIL-Regular.ttf]" at 7pt \zz አረም ናት በኋላ የምትራባው። \end{geez} & \\
%          Transmission & & \\
%          Symptom & & \\
%          Cure & & \\
%     \end{tabular}
%     \caption{Caption}
%     \label{tab:salient_stages}
% \end{table}

% \begin{table}[]
%     \centering
%     \begin{tabular}{c|c|c}
%         Category & YouTube Salient Words & TikTok Salient Words \\
%          Communicable &  & \\
%          Non-Communicable & & \\
%          Women and Children & & \\
%          Reproductive Health & & \\
%          Mental Health & & \\
%     \end{tabular}
%     \caption{Caption}
%     \label{tab:salient_categoies}
% \end{table}

% Overall, we find that ... 
% Below, we detail the methods we used to prepare queries, extract and analyze our data as well as detail our findings. 

\subsection{RQ2: What is the quality of health information retrieved for low-web data languages?} \label{sec:rq2} 
% The goal of \textbf{RQ2}, is to understand who is creating health information in the two target languages. We first looked at the statistics of channels and accounts that posted videos in our dataset. We then qualitatively analyzed a subset of the videos to get a deeper understanding of the content creators.
% To answer \textbf{RQ2}, we qualitatively analyzed 40 YouTube videos and 10 TikTok videos. 
We identified characteristics in relation to the four axis of the CRAP framework in regards to the channels and accounts that post content returned for health related queries in the two low-web data languages: (1) \underline{C}urrency: while content creators post frequently, they do not always post health related content; (2) \underline{R}eliability: content creators rarely cite sources for their claims; (3) \underline{A}uthority: content creators were mainly individual medical doctors or religious leaders; (4) \underline{P}urpose: some channels used their content to drive traffic to their business. Below, we go into detail of our findings for each of the four axis of the CRAP framework. 

\paragraph{Currency:} Currency is a
measure of the recency of the information.
On both platforms, we find that most of the accounts and channels post frequently. However, not all the posts in a given channel are health-related. For instance, one channel, which posted on average 5 times a month in the past 9 months, posted vlogs about their personal life in addition to medical content. 
Across both languages, we find that most health-related videos returned (n=15) were from 2022, with one video from 2024. The oldest video returned for YouTube was from 2018, which was a video on handling stress by a religious leader. For TikTok, most of the videos (n=4) were from December 2024. 
% For YouTube, we find that the date range for the videos is from 
% % \textcolor{red}{(DW: don't abbreviate months)} 
% September 8, 2018, to August 26, 2024. We also find that channels post frequently, with the latest video being uploaded in January 2025. The oldest updated channel was updated on April 2, 2024. 
% We find that most channels are active, posting multiple times a month or a week. 
% \textcolor{red}{DW: this section could be a bit more robust. For each topic, are there, on average, more recent videos available for each topic in each language? Are there certain topics with videos that are older compared to other topics?--- Updated DW: Concerns still remain here. First, the date range makes it difficult to understand how recent this was. For example, maybe mentioning that on average results contained videos that were posted within the last three months with the oldest recommended video having a publication date of more than 13 months. A sub-component of currency is also relevance. Did any of the topics require more updated information and were the sources sensitive and responsive to that? For example, if there were any major breakthroughs on cures or major new trends in symptoms.  }
% \hellina{I could not see a particular pattern since we are qualitatively looking at 50 videos in total that are the most viewed videos. I am not sure if we can generalize from that?}
% For TikTok, we find that the date range of the videos is from December 10, 2024 until October 31, 2024.

\paragraph{Reliability}
% \textcolor{red}{*TODO: add filler sentence summarizing the goal or purpose of this dimension of the framework*}. 
tells us to what degree the information is based on fact versus opinion. 
% For YouTube, we find that some of the videos returned are out of context or are in a language other than the language of the search query. For Tigrinya, we found most of the videos to be in languages other than Tigrinya. A few of the videos were out of context, for instance, some of the videos were posts about lifestyle in Portugal or Vietnam. The remaining videos, while they were about health, were posted in Amharic. 
% In Table \ref{tab:channels} and Table \ref{tab:channels_tiktok}, we give examples of channels we analyzed.
Overall, while the videos in our analysis were fact-based--most were educational videos that list signs and symptoms or give recommendations for natural remedies--a majority of the videos did not cite their sources.  
We find that the videos varied widely in terms of the topics they covered, including educational content about medical conditions, personal stories about experience with health conditions, news reportings, and religious content. 
% Most videos were educational videos, for instance, listing the signs and symptoms of diseases; 
However, most of the videos do not cite their sources--only two YouTube videos cite their sources while none of the TikTok videos cite their sources. Some videos use images from other sources or add visual demonstrations. 
Most videos also gave nutritional advice—i.e. what food to eat for a particular disease. 
For some videos, it was hard to verify the reliability as they were (1) in another language, (2) did not contain medical information; for instance, a query about pregnancy returned a comedy skit about pregnancy cravings, or (3) was religious content; we find that all but three queries related to mental health returned religious content as the top-viewed video
\footnote{See Supplementary Material for channel and account examples.}.

\paragraph{Authority}
% \textcolor{red}{*TODO: add filler sentence summarizing the goal or purpose of this dimension of the framework*}.
gives information on who the creators are and what credentials they have.
Most of the channels and accounts in our analysis are run by individuals whose credentials were hard to verify. 
Although many of the content creators in our analysis were individual medical professionals. 
we find that none of the content creators were verified by the platforms as medical content creators. Additionally, channels that disclose their locations all indicate being located outside of Ethiopia or Eritrea. We find that the channels that disclose their location are located in South Africa (n=1), the US (n=3) and Germany (n=1) for Tigrinya, while for Amharic, we find all channels that disclosed their location set their location to the US (n=9). We find that some creators share their office location (Google Maps link) and phone number. Interestingly, even when there is evidence that the content creator is located in Ethiopia, the channel is registered in the US. Creators also link their profiles to other platforms like Twitter/X, TikTok, and Facebook. We were able to verify 5 medical doctors by following the links they provided to other platforms; for instance, one creator had a ``ECFMG Certified Medical Grad’’ tag on their Twitter account. Some creators post their email and Facebook accounts but do not provide any verification. For TikTok, it was difficult to verify the creators as there is no information in their descriptions, and most of the videos were voice-overs. Two channels had their first name, and one channel had a phone number to contact for asking for services. One creator had a Dr. title before their name, although there were no avenues to verify their credentials. 

% For both Tigrinya and Amharic, on the YouTube videos, we mostly find claims from individual medical doctors and religious leaders for videos that are healthcare-related. For Amharic, we also found a pharmacist based in the US.  Creators also link their profiles to other platforms like Twitter/X, TikTok, and Facebook. We were able to verify 5 medical doctors by following the links they provided to other platforms; for instance, one creator had a ``ECFMG Certified Medical Grad’’ tag on their Twitter account. Some creators post their email and Facebook accounts but do not provide any verification. We also found one video that was a news report of a person's experience with breast cancer; the video included an interview with a medical doctor describing symptoms and risk factors for breast cancer. 

\paragraph{Purpose}
% \textcolor{red}{*TODO: add filler sentence summarizing the goal or purpose of this dimension of the framework*}. 
provides insights into whether the information available is pushing a particular agenda. 
% For YouTube, we find that most of the content is fact-based, although it is hard to verify the facts as the creators do not cite their sources.
While most channels were not biased towards a particular agenda, we noted a few videos where the content creators used the video to drive traffic to their business.
In general, most of the videos were about nutrition and foods to eat for a particular disease. 
However, in some cases, the videos included ads or contact information for viewers who were interested in contacting the content creators. 
One YouTube channel and one TikTok account had an ad for a dietary supplement from a company owned by the content creators.
In other videos, we observed that doctors leave their contact information for people who want personal consultations. 
We also found one account that used its TikTok content to derive business for their traditional medical clinic.
Some content creators also add disclaimers in their videos or descriptions that people should seek medical advice when necessary.
The 5 medical doctors we could verify and one of the religious leaders all had a disclaimer that the content was not a substitute for medical diagnosis and that people should talk to their doctors for such services. 
However, fake accounts created using the identities of medical professionals sometimes post scam content.
In one video, one of the content creators we verified as a doctor shared that other people have been using their image and channel name with fake accounts to exploit people, falsely claiming they will send them medication in return for monetary payment.

% Overall, from our analysis using the CRAP framework, we find that 

 \begin{table}[ht!]
     \centering
     \footnotesize
     %\small
     \begin{tabular}{p{1.5cm}|p{3cm}|p{3cm}}
     
     \toprule
         \textbf{Source} & \textbf{Description} & \textbf{Example}\\
         \midrule
         Borrowed Medical Terms  & Some diseases that do not have terms in the local language, and instead use transliterated English words, result in unrelated content being returned. & Searching for `HPV' and `Meningitis' returns results that are about other diseases. \\
        \midrule
         Similar Medical Terms & Some medical terms have similar writings in the local languages & Anemia translates to ``Blood Decline'' in Amharic which is similar to ``Blood Pressure.'' As a result, the videos returned are about ``Blood Pressure.''\\
         \midrule
         Sensitive Medical Conditions & Diseases that are culturally considered to be taboo topics, especially related to reproductive health. & Searching for information on STDs results in videos in Arabic or English\\

         \midrule
         Critical Medical Conditions & Searching for information about chronic medical conditions returns videos on nutritional cures. & Searching for information on ``cancer.'' \\
         \midrule
         Non-Medical Terms & Popular search terms that have nothing to do with the medical content and instead are popular at the time. & Political terms used in hashtags \\
         \bottomrule
     \end{tabular}
     \caption{Five categories of terminology that lead to search results that diverge from the queries.}
     \label{tab:manipulation}
 \end{table}
 \subsection{RQ3: When search queries and results do not align, what are the common characteristics of those results?}

 % \textcolor{red}{@Hellina - see comments about the framing for this RQ}
 
% \textcolor{red}{DW: I think this first sentence would be even stronger if you begin by signaling which analysis helped in extracting these insights. For example, if it was the qual then we could refer to trends in the themes, trends from topic modeling or maybe a combination of both where one provided more nuance.}
%From our qualitative analysis and topic modeling results, 

So far, we found that search results retrieved do not always match their queries, which creates a space for harm to health information seekers---users cannot get the information they are seeking about certain medical conditions. Instead, the content retrieved may be (1) in another language, (2) about an unrelated medical condition, or (3) non-medical content. Further, we find that there are different degrees of human manipulation--i.e, to what extent there is an agent manipulating the health information gap for disingenuous purposes-- in causing the search results to diverge from their queries. In this section, we explore in-depth the divergence of search results from their search terms.

We identified five cases where the search results diverge from their queries. 
  Below, we describe these five cases listed in order of degree of human manipulation for dubious purposes. Starting with the cases where there is no human manipulation, we first present borrowed medical terms and similar medical terms, where the divergence of retrieved results from the queries can be attributed to the \textit{linguistic} limits of the search engine in handling health queries that are in low-web data languages.
  % In these two cases, the lack of data in low-web data languages, and the lack of robust retrieval systems leads to a case where the search engine returns content that is ``close enough.'' However, in a critical setting of healthcare, ``close enough'' may still be harmful. 
  We then present the cases where the manipulation is still largely in the failure of the retrieval algorithm, but may be shaped by socio-cultural facets. Here, we have sensitive medical terms where the divergence may be due to the lack of content about tabooed topics in the \textit{cultural} context of low-web data languages. We then present the case of critical medical conditions where there may be (un)intentional interference with the search experience for critical medical cases that may require authoritative information. Lastly, we present the case where there is active human manipulation--using non-medical unrelated terms as hashtags, content creators manipulate the retrieval algorithm to boost content that is not related to the query. 
% \begin{itemize}
    \paragraph{Borrowed Medical Terms:} Some diseases and health conditions are referred to with their English name or with their scientific name. In some cases, this is because the medical condition does not have a name in the target language--for instance, AIDS is referred to as AIDS in both languages (although, it is written with the Ge'ez characters). In other cases, the medical condition might have a name in the target language but the name might not be widely used or might be used interchangeably with the English or scientific name--e.g ``cancer'' has a local name in both languages but may also be referred to with its English name. In our analysis, we found that some of the search results returned for health conditions that are referred to by borrowed medical terms were unrelated to the queries.  For instance, search queries about ``HPV'' returned videos (1) about computer problems--some related to computer viruses, (2) about different medical problems--like acid reflux, (3) sexual advice that violates the YouTube platform policy.
    
    \paragraph{Similar Medical Terms:} We also found that distinct medical conditions that have similar words are confused in the search results: for instance, for Amharic, the medical term for Anemia is ``\underline{dem} manes'' which literally translates to ``Blood Decline.'' However, the search results for queries in Amharic included videos about ``Blood Pressure'' which in Amharic is ``\underline{dem} bezat.'' Thus, while the videos were medical in nature, they were about an unrelated medical condition than what was searched. 
    
   \paragraph{Sensitive Medical Conditions:} Search results for medical conditions that were sensitive or were about culturally tabooed topics also diverged from their queries. Particularly, searching for STDs resulted in videos in languages other than the language of search. For instance, searching for ``What is the cure for syphilis?'' in Amharic repeatedly returned videos about Syphilis in Arabic and English. On the other hand, searching for ``How can I prevent HIV?'' in Tigrinya consistently returned videos in Amhairc. Language mixing in search results for sensitive medical conditions is also evident in Figure \ref{fig:salient_terms} where the reproductive health category has more salient terms that were different from the language of search. 

   \paragraph{Critical Medical Conditions:} In cases where the health condition is critical--e.g, cancer or diabetes, we find that the search results were more about natural cures. Usually, the video titles promise a quick and lasting fix: for instance, a query about breast cancer cure returned Amharic videos about ``fruit that kills cancer after just a few minutes'' and videos about ``anti-cancer foods'' that will ``eliminate cancer if you eat these 8 foods frequently.''

   \paragraph{Non Medical Terms:} We also found that content creators use non-medical, popular terms as hashtags to boost their content. For instance, a video about symptoms and cures for evil eye was returned for a query about meningitis; the video used hashtags of popular TV shows and names of political figures in its video description, which may have boosted its ranking in the search engine.

% \section{Discussion}
% \textcolor{red}{DW -- Hellina see comments}
% As we have shown in sections \ref{sec:rq1} and \ref{sec:rq2}, search results for health queries in low-web data languages (1) may not be available in the search language, (2) are posted by content creators whose credentials are hard to verify, (3) may include search results unrelated to the query. These three findings connect to the health information gap definition we present in section \ref{sec:background}. With those findings in mind, we answer our \textbf{RQ3} on the consequences of the health information gap.

\section{5 Proposing a Data Horizon}
Recall that a data void is a gap in digital information where search queries do not have high-quality nor high-quantity information, resulting in a \textit{void} that can be manipulated by bad actors. 
In this paper, we investigate a \ourterm--the search experience right before one goes into a data void. As briefly described in Section 1, a \ourterm is the boundary where the information landscape begins to fall apart due to systemic constraints on information access.
The \ourterm in a health information gap for low-web data languages has five axes of resource constraints.
% The \ourterm in a health information gap for low-web data languages has five axes of resource constraints that, in cohesion, result in ``obscure search queries [with] few results associated with them''\cite{golebiewski_data_2019} and, thus, risks the creation of a data void.
% We use this section to describe how the health information seeking experience for low-web data language speakers is largely shaped by 5 interacting axis of resource constraints. 
Below, we describe each of the five axes and connect them to our findings from the previous section to demonstrate how they shape the experiences of low-web data speakers seeking health information online. 
% introduce a new type of data void, one that is the result of the interaction of five distinct but interacting phenomena:  

\paragraph{Lack of Digital Resources in Target Language:} Due to the low-quality web data, the search results for health queries in the two low-web data languages include information in other languages or unrelated information in the target language (Section 4.1). Limits in digital resources available in the language are further reflected in search results for queries with borrowed and similar medical terms, where results returned diverge from the queries (Table \ref{tab:manipulation}). 

\paragraph{Critical Domain Setting:} In a critical domain such as healthcare, lack of verified health information may lead to life-threatening consequences (Section 2). As presented in Section 4, we find that the information returned for health queries may not always be medical content or may be about a health condition unrelated to what was in the query. 
Further, the information returned may not be in the language of search, creating a language disparity in access to health information.
% Additionally, as Section 4.2 shows, it is difficult to verify the credentials of the content creators who post medical content. Further, most of the content is posted by individual content creators as opposed to `authoritative' institutions, making it that much more difficult to accredit authority. 

\paragraph{Lack of Verification Mechanisms:} As we present in Section \ref{sec:rq2}, none of the YouTube channels and TikTok accounts we analyzed were verified by the platforms as authorized medical information providers. Independently verifying their credentials required following links to other platforms. We verified 5 medical doctors on YouTube and could not verify the credibility of any of the creators on TikTok. 

\paragraph{Out-Of-Context Platform Use:} As we have discussed in Section 1, while social media platforms are primarily designed to foster social communication, they are increasingly used as search platforms. As a result, features that are optimized for social engagement may interfere with the quality of search results returned; for instance, we find that hashtags with unrelated, non-medical terms result in videos that are not related to the query being returned as search results (Section 4). 

\paragraph{Shift in Cultural Context:} In Section 4, our findings show that salient terms for health queries in low-web data languages include nutritional advice as well as religious and spiritual terms. Further, from our analysis of the top-viewed videos, we find that videos from religious leaders are among the top-viewed videos for mental health-related queries. Additionally, we find that searching for health information about taboo topics--such as STDs and reproductive health--may result in search results in other languages. These could be the results of the cultural context in which the two low-web data languages are spoken.

The interaction of the five phenomena described above creates a \ourterm. The \ourterm is the stage right before we go into a data void, where the active manipulation has not begun.
As Section 4 shows, the \ourterm is mainly manipulated through (1) algorithmic failures in handling borrowed medical terms, similar medical terms, and queries about sensitive medical conditions, (2) (un)intentionally by unverified content creators who post natural cures for chronic and critical medical conditions, and (3) active manipulation by content creators who use popular but non-medical terms as hashtags to boost their content's ranking in retrieval. 

In Table \ref{tab:interaction}, we show how the five axes of a \ourterm interact with the five sources of search results that diverge from their queries. As discussed in Section 4, the sources of search results that diverge from their queries as listed in order of degree of human manipulation. We find that all five sources of search results that diverge from their queries are affected by the lack of digital resources, the out-of-context use of social media platforms for health information seeking, and the shift in cultural context. For instance, in all five cases, the lack of digital resources in low-web data languages means the algorithm has no documents related to the query. As a result, the retrieval system returns what is ``close enough''--be it in another language or about unrelated content. 
Similarly, the out-of-context use of the social media platforms means that what is returned is not necessarily ranked based on the quality and credibility it has to the query. In the case of non-medical terms, the out-of-context use becomes a breeding ground for manipulation---affordances such as hashtags that were designed for social connection are instead used to manipulate retrieval ranking. 
Further, the shift in cultural context requires algorithms that handle linguistic features such as borrowed medical terminology and similar medical terms. Cultural context is also important in determining what types of non-medical terms might be used in a given language. 

Conversely, the critical domain affects all but non-medical terms in the five sources we identified as non-medical terms can be used to manipulate the algorithm in any domain. 
The lack of verification, however, only affects non-medical terms; even if the content returned for borrowed medical terms or sensitive medical conditions is coming from verified sources, it will still fail in providing the information the user is seeking. Take the case of searching for content about anemia. As we have demonstrated above, the search results for such queries may be about blood pressure due to the similarity of the medical terms in the Amharic language. Even if the content about blood pressure is coming from a verified medical professional, it fails to provide the information that was requested in the query. Conversely, for non-medical terminology, having verification for content creators will allow people to distinguish between what content to trust, regardless of how it is ranked in the search results. 
% can be manipulated 1) by content creators who use non-medical terms to boost their content, 2) (un)intentionally through the spread of unverified natural remedies, even for chronic diseases, and 3) algorithmically due to the inherent limits of information in the low-web data languages. 

 \begin{table*}[]  
     \centering
     \footnotesize
     \begin{tabular}{c|p{1.5cm}|p{1.5cm}|p{1.5cm}|p{1.5cm}|p{2cm}}
     % \small
     \toprule
         \textbf{Source} & \textbf{Digital Resource}  & \textbf{Out-of-Context Use}  & \textbf{Culture Shift } & \textbf{Critical Domain} & \textbf{Lack of Verification}  \\
         \midrule
         Borrowed Medical Terms & \checkmark & \checkmark &  \checkmark &  \checkmark & \\ 
         Similar Medical Terms & \checkmark & \checkmark & \checkmark & \checkmark &  \\
         Sensitive Medical Conditions & \checkmark & \checkmark & \checkmark & \checkmark &  \\
         
         Critical Medical Conditions & \checkmark & \checkmark & \checkmark & \checkmark &  \\

         Non Medical Terms & \checkmark & \checkmark & \checkmark &  & \checkmark \\
         \bottomrule
     \end{tabular}
     \caption{Table showing which axes of a \ourterm influence the sources of search results that diverge from their queries.}
     \label{tab:interaction}
 \end{table*}

\begin{figure}[htp]
    \centering
    \includegraphics[width=8.5cm]{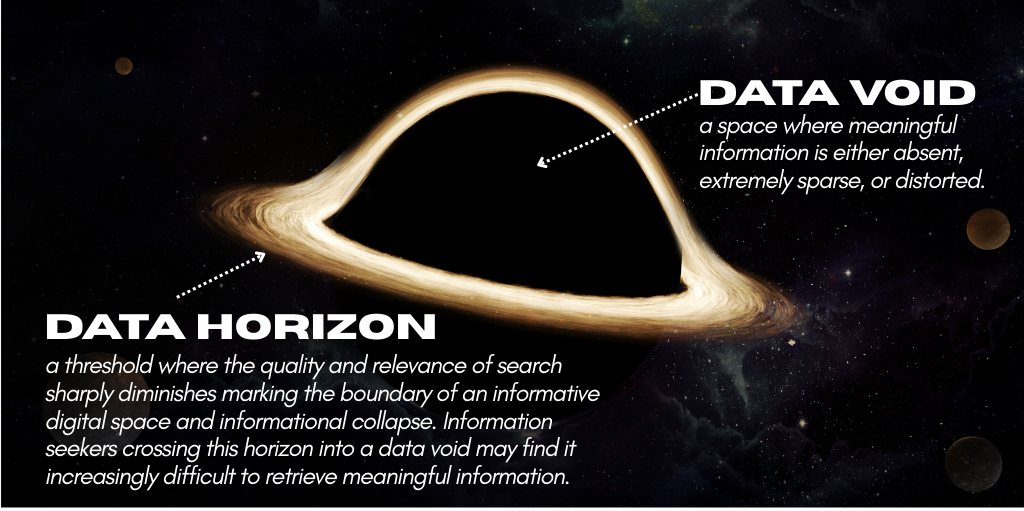}
    \caption{Conceptual illustration of a data void and its data horizon representing the boundary beyond which relevant, credible, or reliable information becomes unavailable}
    \label{fig:horizon} 
\end{figure}

\section{6 Discussion}

% \textcolor{red}{@Hellina: see Figure 3. Consider working into the discussion section and referencing it there. At think at this point its fine to also use language that signals acknowledgment of the terms from astrophysics. For example, describing the data horizon as a point where the fabric of an informative digital space begins to warp, acting like an invisible boundary around the data void. Information seekers who cross this boundary may find it increasingly difficult to retrieve reliable or meaningful information. Then mention how studying the horizon could help in identifying data voids which has been notoriously difficult to identify.  }

As the use of social media platforms like YouTube and TikTok evolved from their intended design of being platforms for social connection to serving as substitutes for search engines, users are at higher risk of exposure to harm~\cite{sylvia_chou_where_2020}. One such harm is due to data voids--information vacuums that are ripe for exploitation by malicious actors~\cite{golebiewski_data_2019}. 

We introduce data horizons, which are areas in the information landscape where the fabric of information is stretched by systemic constraints on access to information, causing it to wrap and act like an invisible boundary around a data void. In Figure \ref{fig:horizon}, we show this phenomenon. At the data horizon, the quality of information is spread thin due to sociotechnical constraints (Section 5). As a result, individuals searching for information while at the horizon may not be able to access the information they seek. However, they are also not always targets of active manipulation--instead, their experiences are largely shaped by the degrading algorithmic performance and the shift in socio-cultural context. 
% Unlike data voids, their saliency is not determined by external factors. 
To understand the characteristics of data horizons and identify how they are formed, we focused on health queries in two low-web data languages, Amharic and Tigrinya. We set out to (1) characterize search results, (2) understand the quality of health information, and (3) investigate causes for search results that diverge from their queries. 

We find that the health information that is available online in the two languages of study leans towards providing nutritional and natural remedies, includes religious content, especially for mental health queries (Section 4.1). Specifically for mental health queries, we find that the most viewed videos are religious content. This indicates a tendency by consumers of health information to seek information that may not be classified as authoritative medical content. However, as prior work shows, the lack of authoritative content about topics that are not mainstream may lead to unverified and at times harmful content reaching consumers~\cite{abebe_using_2019}. We want to emphasize that we are not making claims that religious content is unequivocally invalid for mental health support; many religious societies rely on their spiritual practices for mental health support~\cite{oewel_approaches_2024}. Rather, we argue that knowing what type of content is most effective in reaching this particular community would allow us to target tailored interventions. One of the religious content creators states in their video that they are ``not opposing seeking medical help.'' Hence, there is an opportunity to, for instance, collaborate with religious leaders to create awareness about mental health issues. 

% and may not always be in the language of the search (Section 4.1).  
\citet{golebiewski_data_2019} state, for queries that are associated with authoritative content, such as medical queries, search engines are more likely to return authoritative medical content in their top results. However, medical resources are not equally available in all languages~\cite{hu_natural_2025}. Additionally, it may be more difficult for accredited health providers from certain communities to be labeled ``authoritative'' (Section 2). 
% As a result, low-web data language speakers have limited avenues to access verified health information. 
Our results also show that it is difficult to verify the credentials of content creators who post health information in low-web data languages (Section 4). It was particularly challenging to verify the credentials for TikTok content creators. Further, while content creators who share health information on online platforms in the two languages of study post frequently, they may not always post health-related content (Section 4). This might indicate a relatable angle for social media health information, as has been shown by prior work~\cite{milton_i_2023}. However, as we have described in Section 2, relatability may result in mistrust by users who consume health information on social media platforms. This factor might be confounded by how content creators rarely cite their sources (Section 4). Hence, low-web data language speakers may have limited access to trusted, verified health information on social media platforms.

As for social media platforms, there are a few avenues to protect users from falling into data voids in low-web data language queries. First, while it might be difficult to verify medical content creators from countries that do not have the proper mechanisms set up, platforms can require content creators to provide their own verifications (e.g. what hospital they work at, certifications they might have e.t.c) on their channel descriptions. While this does not serve as a replacement for the proper accreditation for medical professionals verified by the platforms, it will make it easier for users to make their own verifications. Platforms could also consider adding links to websites of recognized organizations like the WHO on videos that are posting medical content, so users have options to search for further information. Platforms could also encourage content creators to add citations and provide additional sources when posting health-related videos, regardless of them being verified by the platform as a medical professional. However, we caution that this should not take the form of punitive measures against existing content creators who are providing medical information with the limited resources available to them.

By definition, the lack of data in low-web data languages creates a void. This void can lead users to harm, particularly in use cases like healthcare. International and local organizations can partner with content creators to fill the void. Such organizations can (1) help speed up the verification of content creators by providing the necessary certifications for medical professionals from these communities, and (2) help content creators produce more content in these languages with the right resources for citations. The organizations can also create their own content in various languages. 

Data voids are difficult to detect in that they require external triggers to surface---by the time a data void becomes a problem, the harm has already been done~\cite{golebiewski_data_2019}. With the concept of data horizons, however, we can identify data voids \textit{before} they are fully exploited. To uncover a data horizon, we need only look at the \textit{systemic constraints} to information access. As we have shown in Section 5, we can rely on the axis of a data horizon to understand how it manifests. Hence, we can strategize about ways to protect users. 

\section{Limitations and Conclusion} \label{sec:conclusion}
Our paper proposes data horizons--boundaries in the information landscape created due to the tension between various systemic constraints to information access. We used the context of health information seeking in low-web data languages to characterize data horizons. We show that by studying socio-technical constraints that lead to degradation of information quality, we can identify data horizons before they become full-fledged data voids. Our study is limited in that we study only two languages. This was due to the human resource limits on how many languages we could cover. Future work could explore extending our study to more languages and mapping the data horizon to a different critical domain (e.g. legal domain). We hope that our work informs stakeholders on potential avenues for identifying data horizons and protecting users before they fall into data voids. 
\bibliographystyle{ACM-Reference-Format}
\bibliography{references}
\appendix
\end{document}